\documentstyle[aps,preprint,eqsecnum,amsfonts]{revtex}
\begin{document}
%
%

\title{Explicit Derivation of New Hyper-Kahler metric}
\author{\bf Moulay Brahim Sedra}
\address{Universit\'e Ibn Tofail, Facult\'e des Sciences,
 D\'epartement de Physique, B.P. 133, 14000 K\'enitra
{\bf Morocco}\\
and\\
Universit\'e Mohammed V, Facult\'e des Sciences, D\'epartement de
Physique,UFR Physique des Hautes Energies, B.P. 1400 Rabat,
{\bf Morocco}\thanks{mailing address}}
\date{\today}
\maketitle
\begin{abstract}                                                      
Using the harmonic superspace techniques in D=2 N=4, we present
 an explicit derivation of a new hyper-Kahler metric associated to the
 Toda like self interaction $H ^{4+}(\omega, u)=(\frac{\xi ^{++}}{\lambda})^{2}\exp(2\lambda \omega)
 $. Some important features are also
 discussed.
\end{abstract}
\newpage
\section{introduction}

Recently, a particular interest has been devoted to the subject of
hyper-Kahler metrics building which is solved in a nice way in the
harmonic superspace~\cite{1,2} It is an important question of hyper-Kahler
geometry, a subject much studied in modern theoretical physics, more
especially in connection with the theory of gravitational instantons,
moduli problems in monopole physics, string theory, and elsewhere\cite{3,4,5}.

An original contribution about hyper-Kahler geometry, where a complete
geometrical characterization of manifolds for which extended supersymmetry
is allowed was given in\cite{3}. By requiring covariantly constant complex
structures for the extended supersymmetry, the authors showed that there
exist two possibilities given by $N=2$ and $N=4$ supersymmetry for which the
Kahler and hyper-Kahler manifolds play a special role respectively. On the
other hand, these hyper-Kahler structures are of great interest as they
are involved recently in the problem of moduli spaces of monopoles. A more
tractable example being the moduli space of BPS magnetic monopoles \cite{6},
which is shown to possess hyper-Kahler structure.
Although the previous achievements, a lot of things remains to do. The
classification of all complete, regular, hyper-Kahler manifolds remains an
open question to this date, and even for some known examples, the
explicit form of the metric has been difficult or impossible to determine
so far.

Recall that there are only few examples that had been solved exactly,
these are the Taub-Nut and the Eguchi-Hanson hyper-Kahler metrics
exhibiting both a U(2)=SU(2)xU(1) isometries.

	In the present work, we propose another steps toward the
classification of metrics of hyper-Kahler geometry and compute explicitly
the metric associated to the harmonic superspace Toda (Liouville) like
self-interaction. Recall that the problem of finding the
metrics amount to eliminating an infinite number of auxiliary fields. We
derive a new metric which can be useful in the sense that the particular
hyper-Kahler geometrical structure that it suspected to describe is
connected to integrable models via the Toda like self interaction.

	We start in section 2 by recalling some general properties of the
hyper-Kahler metrics building program and present in sections 3 and 4 the details
concerning the derivation of the metric. The explicit form of this metric is exposed
in the appendix. We conclude in section 5 and discuss further the framework for 
possible applications. 

\section{Generalities on hyper-Kahler metrics building from H.S}

We start this section by recalling some general results of the
hyper-Kahler metrics building from the harmonic superspace. The
subject of hyper-Kahler metrics building is an interesting problem of
hyper-Kahler geometry that can be solved in a nice way in harmonic
superspace if one know how to solve the following non linear differential
equations on the sphere $ S^{2}$\cite{2}:
 \begin{mathletters}
 \begin{equation}
\partial^{++}q^{+} - \partial^{++}\left[\frac{\partial
V^{4+}}{\partial(\partial^{++}{\bar q}^{+})}\right] +\frac{\partial
V^{4+}}{\partial{\bar q}^{+}}=0,
\end{equation}
\begin{equation}
\partial^{++}\bar q^+ + \partial^{++}\left[\frac{\partial
V^{4+}}{\partial(\partial^{++} q^{+})}\right]-\frac{\partial V^{4+}}
{\partial q^+}=0,
\end{equation}
\end{mathletters}
where $q^{+} =q^{+}$ ($z, \bar z, u^{\pm})$ and its conjugates $\bar q^{+} =
\bar q^{+}$ ($z, \bar z, u^\pm)$  are complex fields defined on $ C\times S^2$
respectively and are parametrized by the local analytic coordinates
$ z,\bar z$ and the harmonic variable $u^\pm$. The symbol $\partial ^{++}
=u^{+i}{\partial\over \partial u^{-i}} $ stands for the so-called harmonic
derivative and $ V^{4+} = V^{4+}(q,u)$ is an interacting potential
depending in general on $q^+$ ,$\bar q^+$, their derivatives and the $
u^\pm$'s. The fields $q^+$ and $\bar q^+$ are globally defined on the
sphere $ S^2 =SU(2)/U(1)$  and may be expanded into an infinite series in
 power of harmonic variables(for the bosonic part) preserving the total
 U(1) charge in each term of the expansion as given here below
\begin{mathletters}
\begin{equation}
q^{+} (z, \bar z, u) =u^{+}_{i}f^{i}(z,\bar z)
+u^{+}_{i}u^{+}_{j}u^{-}_{k} f^{(ijk)}(z, \bar z)+\cdots
\end{equation}
\end{mathletters}
Note here that Eqs.(2.1), which fix the $u$-dependence of the $q^{+}$'s is in
 fact the pure bosonic projection of a two dimensional $N=4$ supersymmetric HS
superfield equation of motion. The remaining equations carry the spinor
contributions and are shown to describe among others, the space time
dynamics of the physical degrees of freedom namely, the four
bosons $f^{i}(z, \bar z)$, $\bar f^{i}(z, \bar z), i=1,2$, and their $D=2$ $N=4$
supersymmetric partners.

An equivalent approach to write Eq.(2.1) is to use the How-Stelle-Townsend
(HST) realization \cite{7}
\begin{mathletters}
\begin{equation}
{\partial^{++}}^2\omega - \partial^{++}[\frac{\partial H^{4+}}
{\partial(\partial^{++}\omega)}]+\frac{\partial H^{4+}}{\partial \omega}=0,
\end{equation}
\end{mathletters}
 where $\omega=\omega (z, \bar z, u)$ is a real field with zero $U(1)$
 charge defined on ${\Bbb C}\times S^2$ and with its
leading terms of its harmonic expansion given by
 \begin{equation}
 \omega (z, \bar z, u) = u^{+}_{i} u^{-}_{j}f^{ij}(z, \bar z) +
 u^{+}_{i} u^{+}_{j} u^{-}_{k}u^{-}_{l}g^{(ijkl)}(z, \bar z)+\cdots
 \end{equation}
Similarly as in Eq.(2.1), the interaction potential $H^{4+}$ depends
 in general on $\omega$ , its derivatives and the harmonics.
Note the important observation of\cite{8}, that one can always pass
from the $q^{+}$ hypermultiplet to the $\omega$ hypermultiplet via a duality
transformation\cite{9} by making a change of variables.

In the remarkable case where the potentials $H^+$ and $V^{4+}$
do not depend on the derivatives of the fields  $q^+$ and $\omega$,
 Eqs.(1) and (3) reduce then to
\begin{mathletters}
\begin{equation}
\partial ^{++}q^{+} +\frac{\partial V^{4+}}{\partial \bar q^{+}}=0
\end{equation}
\begin{equation}
{\partial ^{++}}^{2}\omega +\frac{\partial H^{4+}}{\partial \omega}=0
\end{equation}
\end{mathletters}
As the solutions of these equations depend naturally on the potentials
$V^{4+}$ and $H^{4+}$, the finding of these solutions is not an easy
question. There are only few examples that had been solved exactly. The
first example is the Taub-Nut metric of the four-dimensional Euclidean
gravity. The potential $V^{4+}$ of this model is given by
\begin{equation}
V^{4+}(q^{+},{\bar q}^+) =\frac{\lambda}{2}{q^{+} {\bar q}^+}^2
\end{equation}
where $\lambda$ is a real coupling constant. For this potential, the
equation of motion reads
\begin{equation}
\partial ^{+}q^{+}+\lambda{q^{+}(\bar q)^{+}}{\bar q}^{+}=0
\end{equation}
and its solution is given by
\begin{equation}
q^{+} (z, \bar z, u) =u^{+}_{i}f^{i}(z,\bar z)\exp(-\lambda
u^{+}_{k}u^{-}_{l}f^{(k}f^{l)})
\end{equation}
Note that the knowledge of this solution is an important steps in the
identification of the metric of the manifold parametrized by the bosonic
fields $ f^{i}(z, \bar z)$ and  $\bar f^{i}(z, \bar z)$ of the $D=2$
$N=4$ supersymmetric non linear Taub-Nut $\sigma$-model.

The second example we consider, is the Eguchi-Hanson model. This model
had been also solved exactly and correspond to the following potential
\begin{equation}
H^{4+} (\omega) =(\frac{\xi ^{++}}{\omega})^{2}
\end{equation}
where the dimensionless quantity $\xi ^{++}$ is given by
\begin{equation}
\xi ^{++}=\xi ^{ij}u^{+}_{i}u^{+}_{j}
\end{equation}
in terms of the real isovector coupling constant $\xi ^{ij}$. Thus
unlike the TN action, the EH action contains explicit harmonics. Details
concerning these two models can be found in the following refs.\cite{2}.

Recently, a new integrable model had been proposed\cite{10}. This model was
obtained by focusing on Eq.(2.5b) and looking for potentials leading to
exact solutions of this equation. The method used in this issue consist
in wondering plausible integrable equations by proceeding as formal
analogy with the known integrable two-dimensional non linear
differential equations especially the Liouville equation and its Toda
generalizations\cite{11}. The important result in this sense was the proposition
of the following potential
\begin{equation}
H ^{4+}(\omega, u)=(\frac{\xi ^{++}}{\lambda})^{2}\exp(2\lambda \omega)
\end{equation}
which leads via Eq.(2.5b) to the following non linear differential
equation of motion
\begin{equation}
\lambda (\partial ^{++})^{2}\omega-(\xi ^{++})^{2}\exp(2\lambda \omega)
\end{equation}
Using the formal analogy with the $SU(2)$ Liouville($SU(N)$ Toda) equation,
we showed that this equation (2.12) is integrable. The explicit solution of this non
linear differential equation reads
\begin{equation}
(\xi ^{++})^{2}\exp(\lambda \omega)=\frac{u^{+}_{i}u^{+}_{j}f^{ij}(z, \bar
z)}{1-u^{+}_{k}u^{-}_{l}f^{kl}(z, \bar z)}
\end{equation}
Furthermore, the origin of the integrability in Eq.(2.12) is shown to
deal with the existence of a symmetry (conformal symmetry) generated by
the following conserved current
\begin{equation}
T^{4+} = {\partial ^{++}}^{2}-\frac{1}{\lambda}{\partial ^{++}}^{2}\omega
\end{equation}
with $\partial ^{++}T^{4+}=0$

\section{Computation of the bosonic metric}
We focus in this section to apply the general method presented in ref.\cite{2}
to derive the hyper-Kahler metric associated to the proposed potential
Eq.(2.11) using the harmonic superspace approach. This method consist in
writing the action describing the general coupling of the analytic
superfield $\Omega$ we are interested in and in deriving the
corresponding equation of motion. For this action which correspond to
some hyper-Kahler manifold, one has only to expand the equations of
motion in spinor coordinates $\theta$, omit the fermions and solve the
auxiliary fields equations. Substituting the solution into the original
action and integrating over the harmonics ones $u^{\pm}$, yields the
required component form of the action from which the hyper-Kahler metric
can be obtained.
Let us first recall that the harmonic superspace (HS) is parametrized
by the supercoordinates $Z^{M}=(z^{M}_{A}, \theta ^{-}_{r}, \bar \theta
^{-}_{r})$ where $z^{M}_{A}=(z, \bar z, \theta ^{+}_{r}, {\bar
\theta}^{+}_{r}, u^{\pm}) $ are the supercoordinates of of the so-called
analytic subspace in which $D=2$ $N=4$ supersymmetric theories are
formulated. The integral measure of the harmonic superspace is given in
the $Z^{M}_{A}$ basis by $d^{2}zd^{4}\theta ^{+}du$. The matter
superfield $(O^{4}, (\frac {1}{2})^{4})$ is realized by two dual
analytic superfields $Q^{+}$ =$Q^{+}(z, \bar z, \theta ^{+}, (\bar
\theta)^{+}, u)$ and $\Omega =\Omega(z, \bar z, \theta ^{+}, (\bar
\theta)^{+}, u)$ whose leading bosonic fields are respectively given by
$q^{+}$ and $\omega$ Eqs.(2.2a) and (2.4). The model we are interested in and
which describe the coupling of the analytic superfield $\Omega$ is given
by the following action\cite{10}
\begin{equation}
S[\Omega ]=\int d^{2}z d^{4}\theta ^{+}du\{\frac{1}{2}(D^{++}\Omega )^{2}
+\frac{1}{2}(\frac{\xi ^{++}}{\lambda })^{2}e^{2\lambda \Omega} \}
\end{equation}
where $\lambda$ is the coupling constant of the model and $\xi ^{++}=
u^{+}_{i}u^{+}_{j}\xi ^{(ij)} $ a constant isotriplet similar to that
appearing in the Eguchi-Hanson model and where $D^{++}$ is the harmonic
derivative given by
\begin{equation}
D^{++}=\partial ^{++}-2 \bar \theta^{+}_{r}\theta^{+}_{r}\partial
_{-2r}
\end{equation}
The equation of motion corresponding to this action Eq.(3.1) reads
\begin{equation}
\lambda{ D^{++}}^{2}\Omega -{\xi ^{++}}^{2}e^{2\lambda \Omega}=0
\end{equation}
where $\Omega$ is the analytic superfield which expand in $
\theta^{+}_{r}$ and $ {\bar \theta}^{+}_{r}$ series as
\begin{eqnarray}
\Omega
&=& \omega+[\theta^{+}_{-}\theta^{+}_{+}F^{--}+\bar \theta^{+}_{+}\bar
\theta^{+}_{-}\bar F^{--}]\nonumber\\
& &+[ \bar \theta^{+}_{-} \theta^{+}_{+}G^{--}+
 \bar \theta ^{+}_{+} \theta ^{+}_{-} \bar G^{--}]\nonumber\\
& &+[ \bar \theta^{+}_{-} \theta^{+}_{-}B^{--}_{++}+
 \bar \theta ^{+}_{+} \theta^{+}_{+}B^{--}_{--}]\nonumber\\
& &+[ \bar \theta ^{+}_{-} \theta ^{+}_{-} \bar
\theta ^{+}_{+} \theta ^{+}_{+}\Delta ^{ -4 }]
\end{eqnarray}

Substituting Eq.(3.4) into Eq.(3.3), one obtains the following non
linear differential equations

\begin{equation}
\lambda {\partial ^{++}}^{2}\omega-{\xi^{++}}^{2}e^{2\lambda\omega}=0\\
\end{equation}

 \begin{mathletters}
 \begin{equation}
 {\partial ^{++}}^{2}F^{--}-2{\xi^{++}}^{2}F^{--}e^{2\lambda \omega}=0\\
\end{equation}
 \begin{equation}
 {\partial ^{++}}^{2}{\bar F}^{--}-2{\xi^{++}}^{2}{\bar F}^{--}e^{2 \lambda \omega}=0\\
\end{equation}
\end{mathletters}

 \begin{mathletters}
 \begin{equation}
 {\partial ^{++}}^{2}G^{--}-2{\xi^{++}}^{2}G^{--}e^{2\lambda\omega}=0\\
\end{equation}
\begin{equation}
 {\partial ^{++}}^{2}{\bar G}^{--}-2{\xi^{++}}^{2}{\bar G}^{--}
 e^{2\lambda\omega}=0\\
\end{equation}
\end{mathletters}

 \begin{mathletters}
 \begin{equation}
 {\partial ^{++}}^{2}B^{--}_{++}-2{\xi^{++}}^{2}B^{--}_{++}e^{2\lambda\omega}
  =4\partial^{++}\partial_{++}\omega\\
\end{equation}
 \begin{equation}
 {\partial^{++}}^{2}B^{--}_{--}-2{\xi^{++}}^{2}B^{--}_{--}e^{2\lambda
 \omega}=4\partial^{++}\partial_{--}\omega
 \end{equation}
 \end{mathletters}

\begin{eqnarray}
{\partial^{++}}^{2}\Delta^{(-4)}&-&4\partial^{++}\partial_{--}B^{--}_{++}
-4\partial^{++}\partial_{++}B^{--}_{--}\nonumber\\
& &\ \ \ \ \ = 2{\xi^{++}}^{2}e^{2\lambda\omega}[\Delta^{(-4)}+2\lambda (F^{--}{\bar
F}^{--}-G^{--}{\bar G}^{--}+B
^{--}_{++} B^{--}_{--})]
\end{eqnarray}

The Liouville-like equation of motion Eq.(3.5), is a constraint equation
fixing the dependence of $\omega$ in terms of the physical bosonic
fields $f^{(ij)}$ of the $D=2$ $N=4$ hypermultiplet. The knowledge of the
solution of this nonlinear equation is necessary as it is one of the main crucial steps in
this program. The second set of relations Eqs.(3.6-8) describe the
equations of motion of the auxiliary fields $F^{--}$, $G^{--}$ and
$B^{--}_{rr}$ of canonical dimension one.The last equation Eq.(3.9) gives the
equation of motion of the Lagrange field $\Delta^{(-4)}$ of canonical dimension two in
terms of $\omega $ and the other auxiliary fields. To solve these equations of
motion, one start first by solving the Liouville-like equation of motion
Eq.(3.5) whose solution[11], originated from integrability and conformal symmetry
in two dimensions, reads in the HS language  as\cite{10}
\begin{equation}
\xi^{++}e^{\lambda\omega}=\frac{f^{++}}{1-f}
\end{equation}
Next, we integrate the action Eq.(3.1) with respect to the Grassmann
variables $\theta $. One find the following result

\begin{eqnarray}
S=\int d^{2}zdu&\{&[\partial^{++}\omega\partial^{++}\Delta^{(-4)}+
   \frac{1}{\lambda}{\xi^{++}}^{2}e^{2\lambda\omega}\Delta^{(-4)}]\nonumber\\
& + & [\partial^{++}F^{--}\partial^{++}{\bar F}^{--}+2{\xi^{++}}^{2}
   e^{2\lambda\omega}F^{--}{\bar F}^{--}]\nonumber\\
& - & [\partial^{++}G^{--}\partial^{++}{\bar G}^{--}+2{\xi^{++}}^{2}
   e^{2\lambda\omega}G^{--}{\bar G}^{--}]\nonumber\\
& + & [\partial^{++}{B^{--}_{--}}\partial^{++}{B^{--}_{++}}-2\partial^{++}
   \omega(\partial_{--}{B^{--}_{++}}+\partial_{++}{B^{--}_{--}})]\nonumber\\
& + &
[\partial^{++}B^{--}_{--}\partial^{++}B^{--}_{++}-2(\partial_{++}
   \omega\partial^{++}B^{--}_{--}+\partial_{--}\omega\partial^{++}B^{--}_{++})\nonumber\\
   & & \ \ \ \ \ +2{\xi^{++}}^{2}e^{2\lambda\omega}B^{--}_{--}B^{--}_{++}]\}
\end{eqnarray}

Using the equation of motion for $\omega$ Eq.(3.5), one show easily that
the following harmonic integrals are vanishing
\begin{eqnarray}
&\int & du[\partial^{++}\omega\partial^{++}\Delta^{(-4)}+
   \frac{1}{\lambda}{\xi^{++}}^{2}e^{2\lambda\omega}\Delta^{(-4)}]=0\\
 & &\int du[\partial^{++}F^{--}\partial^{++}{\bar F}^{--}+2{\xi^{++}}^{2}
   e^{2\lambda\omega}F^{--}{\bar F}^{--}]=0\\
& &\int du[\partial^{++}G^{--}\partial^{++}{\bar G}^{--}+2{\xi^{++}}^{2}
   e^{2\lambda\omega}G^{--}{\bar G}^{--}]=0\\
& &\int du[\partial^{++}B^{--}_{--}\partial^{++}B^{--}_{++}-2(\partial_{++}
   \omega\partial^{++}B^{--}_{--}+\partial_{--}\omega\partial^{++}B^{--}_{++})\\
& & \ \ \ \ +2{\xi^{++}}^{2}e^{2\lambda\omega}B^{--}_{--}B^{--}_{++}]=0
\end{eqnarray}
These vanishing integrals serve to eliminate the auxiliary fields $ F$,$ G$ and
$\Delta$. The resulting bosonic action is then
\begin{eqnarray}
S=&\int & d^{2}zdu\{4\partial_{++}\omega\partial^{++}B^{--}_{--}-2\partial^{++}
\omega(\partial_{--}B^{--}_{++}+\partial_{++}B^{--}_{--})\nonumber\\
   & & \ \ \ \ \ -2\lambda{\partial^{++}}^{2}\omega B^{--}_{--}B^{--}_{++}\}
\end{eqnarray}

In order to obtain the purely bosonic theory, one have to reduce more
this action which depends now only on the fields $B$ and $\omega$. To do
this, one needs to solve the differential equation Eq.(3.8)
for the auxiliary fields $B^{--}_{rr}, r=\pm$,namely

\begin{equation}
{\partial^{++}}^{2}B^{--}_{rr}=2{\xi^{++}}^{2}e^{2\lambda\omega}B^{--}_{rr
}+4\partial^{++}\partial_{rr}\omega  ,
\end{equation}

this equation gives the relation between $B^{--}_{rr}$ and the fields
$\omega$. Finding a solution for this equation is not an easy exercise,
because the behavior of $B^{--}_{rr}$ seems to be non linear in term of
$\omega$. But, remarking just the fact that when $\omega$ is zero or a
space-time constant, a particular solution of $B^{--}_{rr}$ is zero. One
propose then a solution of Eq.(3.18) of the form:

\begin{equation}
B^{--}_{rr}=\xi^{--}\partial_{rr}\omega  ,
\end{equation}

where $\xi^{--}=u^{-}_{i}u^{-}_{j}\xi^{ij}$ is an $SU(2)$ triplet constant
for which one can set $\partial^{++}\xi^{--}=2$. Injecting this solution
into the action Eq.(3.17) one obtains

\begin{eqnarray}
S=&\int &d^{2}zdu\{\partial_{++}\partial_{--}\omega
[-2\lambda{\xi^{--}}^{2}{\partial^{++}}^{2}\omega +4\lambda\xi^{--}\partial^{++}\omega+8]\nonumber\\
& & \ \ \ \ \
-4\xi^{--}\partial^{++}\omega\partial_{++}\partial_{--}\omega\}
\end{eqnarray}
showing the dependence of the action only on the bosonic degrees of
freedom $f=u^{+}_{i}u^{-}_{j}f^{ij}$ defined in Eq.(3.10). What remains to do now is to use the
equation of motion Eq.(3.10) for $\omega$ and integrate over the
harmonics $u$ to derive the purely bosonic action from which one can
easily identify the metric associated to our hyper-Kahler potential
\begin{equation}
H ^{4+}(\Omega, u)=(\frac{\xi ^{++}}{\lambda})^{2}\exp(2\lambda \Omega)
\end{equation}

\section{Finding the metric}
We start from the action Eq.(3.20) and consider the solution
Eq.(3.10) which imply
\begin{equation}
\partial_{rr}\omega=\frac{1}{\lambda}[\frac{\partial_{rr}f^{++}}{f^{++}}
+\frac{\partial_{rr}f}{1-f}]
\end{equation}
Injecting this formal expression into the bosonic action Eq.(3.20), one
obtains the following result
\begin{eqnarray}
S[f]=\int d^{2}zdu\frac{1}{\lambda^2}\{&8&(\frac{\partial_{--}f^{++}
\partial_{++}f^{++}}{(f^{++})^2}+\frac{\partial_{++}f^{++}\partial_{--}f}
{f^{++}(1-f)}\nonumber\\
&+&\frac{\partial_{--}f^{++}\partial_{++}f}{f^{++}(1-f)}+
\frac{\partial_{--}f\partial_{++}f}{(1-f)^2})\nonumber\\
& \ \ + \ \ &4\xi^{--}(2\frac{\partial_{--}f^{++}\partial_{++}f^{++}}{f^{++}(1-f)}-
\frac{\partial_{--}\partial_{++}f^{++}}{1-f}\nonumber\\
&+&\frac{\partial_{++}f^{++}\partial_{--}f+
\partial_{++}f\partial_{--}f^{++}+ f^{++}\partial_{++}
\partial_{--}f}{(1-f)^{2}})\nonumber\\
&-&2{\xi^{--}}^{2}(\frac{\partial_{--}f^{++}\partial_{++}f^{++}}{(1-f)^2}+
f^{++}[\frac{\partial_{--}f\partial_{++}f^{++}}{(1-f)^3}+
\frac{\partial_{++}f\partial_{--}f^{++}}{(1-f)^3}]\nonumber\\
& \ \ + \ \ &{f^{++}}^2\frac{\partial_{++}f\partial_{--}f}{(1-f)^4})\}\nonumber\\
\end{eqnarray}
where
\begin{mathletters}
\begin{equation}
f=u^{+}_{(i}u^{-}{j)}f^{(ij)}+f^0\nonumber\\
\end{equation}
\begin{equation}
f^{++}=u^{+}_{(i}u^{+}_{j)}f^{(ij)}\\
\end{equation}
\begin{equation}
\xi^{--}=u^{-}_{(i}u^{+}_{j)}\xi^{(ij)}\nonumber\\
\end{equation}
\end{mathletters}
and where the symbol $(ij)$ stands for the symmetric part of the tensor
indices $i,j$. The antisymmetric part $f^{++}$ is omitted as it leads
simply to constant terms.
Note that the obtained action Eq.(4.2) contains terms with
singularity around $f=u^{+i}u^{-}_{i}$. This singularity is originated
from the solution of Toda-(Liouville)like equation of motion Eq.(3.10).
As it is worth stressing that not every solution of the continual
Toda(Liouville) equation of motion has a good space time interpretation,
since the corresponding metric might be incomplete with singularities,
it seems at first sight that our hyper-Kahler metric will be incomplete.
But using some algebraic manipulations, one can derive the complete form
of the metric inspired from the first leading terms in a nice way.
To derive the metric from the bosonic action Eq.(4.2), one use steps by
steps the following operations. Starting once again from Eq.(4.2) and
considering the following approximation
\begin{equation}
\frac{1}{(1-f)^\epsilon}=\sum^{\infty}_{i=0}\frac{(\epsilon +i
-1)!}{(\epsilon-1)!i!}f^{i}, \epsilon =1, 2, 3\cdots
\end{equation}
with $f=u^{+}_{(i}u^{-}{j)}f^{(ij)}+f^0$. Injecting this expression into
Eq.(4.2) and integrating over the harmonics once the power $f^{i}(z,
\bar z)$ of the bosonic field $f$ in Eq.(4.4) are expressed as  a
series in terms of $f^{(ij)}, f^0 $ and the symmetrized product of
harmonics. To do this, one have also to use the standard reduction
identities\cite{2}
\begin{mathletters}
\begin{equation}
u^{+}_{i}u^{+}_{(j_{1}}\cdots u^{+}_{j_{n}}u^{-}_{k_{1}}\cdots
u^{-}_{k_{m)}}=u^{+}_{i}u^{+}_{(j_{1}}\cdots u^{-}_{k_{m)}}+
\frac{m}{m+n+1}\varepsilon_{i(k_{1}}u^{+}_{(j_{1}}\cdots u^{+}_{j_{n}}
u^{-}_{k_{2}}\cdots u^{-}_{k_{m)}}
\end{equation}
\begin{equation}
u^{-}_{i}u^{+}_{(j_{1}}\cdots u^{+}_{j_{n}}u^{-}_{k_{1}}\cdots
u^{-}_{k_{m)}}=u^{-}_{i}u^{+}_{(j_{1}}\cdots u^{-}_{k_{m)}}-
\frac{n}{m+n+1}\varepsilon_{i(j_{1}}u^{+}_{(j_{2}}\cdots u^{+}_{j_{n}}
u^{-}_{k_{2}}\cdots u^{-}_{k_{m)}}
\end{equation}
\end{mathletters}
and the $u^{\pm}_{i}$ integration rules
\begin{equation}
$$ \int du (u^{+})^{(m}(u^{-})^{n)}(u^{+})_{(k}(u^{-})_{l)}=\left\{
\begin{array}{ll}
\frac{(-1)^{n}m!n!}{(m+n+1)!}\delta^{(i_{1}}_{(j_{1}}\cdots
\delta^{i_{m+n)}}_{j_{k+l)}}        & {\rm if\ } {m=l \ \ \ and  \ \ \ n=k}\\
0 & {\rm  otherwise}
\end{array}
\right.$$
\end{equation}
with
\begin{equation}
(u^{+})^{(m}(u^{-})^{n)}\equiv u^{+(i_{1}}\cdots u^{+i_{m}}u^{-j_{1}}u^{-j_{n)}}
\end{equation}
To finally integrate over the harmonics, one learn from the content of
the action Eq.(4.2) in $ u^{\pm}$, that we should use the approximation
Eq.(4.4) at least up to the six order in $\varepsilon$. Lengthy and
very hard calculations leads finally to the following purely bosonic
action
\begin{eqnarray}
S_{bos}[f]=\int d^{2}z\frac{1}{\lambda^2}&\{&A_{ijkl}\partial_{++}f^{ij}
\partial_{--}f^{kl}\nonumber\\
&+&B_{ij}[\partial_{++}f^{0}\partial_{--}f^{ij}+
\partial_{++}f^{ij}\partial_{--}f^{0}]\nonumber\\
&+&C_{ij}\partial_{++}\partial_{--}f^{ij}
+ D\partial_{++}\partial_{--}f^{0}\nonumber\\
&+&E\partial_{++}f^{0}\partial_{--}f^{0}
+F\partial_{++}f^{ij}\partial_{--}f^{ij}\}
\end{eqnarray}
from which one can easily derive the metric. The tensor components of
this metric are $A_{ijkl}, B_{ij}, C_{ij}, D, E$, and $ F$ such that
\begin{eqnarray}
A_{ijkl}=A_{jikl}=A_{ijlk}\nonumber\\
B_{ij}= B_{ji}\\
C_{ij}= C_{ji}\nonumber\\
\end{eqnarray}
The first explicit expression obtained for this metric is of course
incomplete due the previous approximation. The missing terms in this
metric are easily recuperated by looking just at the behavior of the
first leading terms of the components $A_{ijkl}, B_{ij}, C_{ij}, D, E$,
and $ F$ . We present in the appendix, the complete expression of the
metric described by the bosonic field $f$.
From the purely bosonic action Eq.(4.8), one can read lot of properties.
Note for the moment only the following. The constant $\lambda$ is just the coupling
constant of the Liouville- (Toda)like theory . The components of the
bosonic (physical) field $f$ are dimensionless and are interpreted as
the internal coordinates of the hyper-Kahler manifold associated to the
proposed potential Eq.(2.11). Since the metric $(A_{ijkl}, B_{ij},
C_{ij}, D, E, F)$ is dimensionless and does not explicitly depend on
the coupling $\lambda$, one have to pass from the $f=(f^0, f^{ij})$
to the true physical bosonic field by performing the
following rescaling $ f=\lambda f_{phys.}$

\newpage
\section {conclusion}
We have derived explicitly the metric associated to the $SU(2)$
Toda-like hyper-Kahler Potential. The idea to work this metric was first
introduced in\cite{10}. But only recently when studying the Witten-article\cite{12}
(where the author explain that one of the reason to go to eleven
dimensional M-theory is that the study of sixbranes, described by the
multi-Taub-Nut hyper-Kahler metric,  become more simple)we asked the question
what will be the importance of our metric in this context?
This and others important questions  which can
help to achieve the old problem of classifying all the Hyper-Kahler
manifolds  will be discussed in future works.

\section* {Acknowledgement} 
I would like to thank I. Bandos, K.S. Narain, M. O'loughlin and
E.H. Saidi for valuable discussions. 
I would like to thank Professor M. Virasoro for good hospitality
at ICTP where this work was done. I acknowledge also  Professor S. Randjbar-Daemi
and the High Energy section for scientific help.

\newpage 
\appendix 
\section*{} 
\begin{eqnarray*} 
A_{ijkl}
&=&\frac{16}{\lambda^2}\frac{1}{f^{(ij)}f^{(kl)}}
   -\frac{168}{\lambda^2}\xi^{eh}\xi^{mn}f^{qr}f^{st}
   f_{(eh}f_{mn}f_{qr}f_{st}f_{ij}f_{kl)}\\
& &+\frac{8}{30\lambda^2}\sum_{N=0}^{\infty}\left\{\frac{(N+3)!}{2N!}
   {f^0}^N+\sum_{n=1}A_{1}(N,n){f^0}^{N-2n}(ff)^n\right\}f_{(ij}f_{kl)}\\
& &-\frac{4}{3\lambda^2}\sum_{N=0}^{\infty}\left\{(N+1)
   {f^0}^N+\sum_{n=1}A_{2}(N,n){f^0}^{N-2n}(ff)^n\right\}
   \left(\frac{f_{(ij)}}{f^{(kl)}}+\frac{f_{(kl)}}{f^{(ij)}}\right)\\
& &-\frac{14}{15\lambda^2}\sum_{N=0}^{\infty}\left\{\frac{(N+2)!}{2N!}
   {f^0}^N+\sum_{n=1}A_{3}(N,n){f^0}^{N-2n}(ff)^n\right\}\\
& &\times\xi^{mn}\left(\frac{f_{(ij}f_{mn)}}{f^{(kl)}}+\frac{f_{(mn}f_{kl)}}
   {f^{(ij)}}\right)\\
& &+\frac{16}{3\lambda^2}\sum_{N=0}^{\infty}\left\{{f^0}^N+
   \sum_{n=1}A_{4}(N,n){f^0}^{N-2n}(ff)^n\right\}
   \left(\frac{\xi_{(ij)}}{f^{(kl)}}+\frac{\xi_{(kl)}}{f^{(ij)}}\right)\\
& &-\frac{2}{35\lambda^2}\sum_{N=0}^{\infty}\left\{\frac{(N+4)!}{6N!}
   {f^0}^N+\sum_{n=1}A_{5}(N,n){f^0}^{N-2n}(ff)^n\right\}\\
& &\times\xi^{mn}f_{(mn}f_{ij}f_{kl)}\\
& &-\frac{4}{9\lambda^2}\sum_{N=0}^{\infty}\left\{(N+1)(N+2)
   {f^0}^N+\sum_{n=1}A_{6}(N,n){f^0}^{N-2n}(ff)^n\right\}(\xi_{(ij)}f_{(kl)}
   +\xi_{(kl)}f_{(ij)})\\
& &-\frac{1}{63\lambda^2}\xi^{mn}\xi^{rs}f_{(ij}f_{kl}f_{mn}f_{rs)}\\
& &-\frac{2}{45\lambda^2}\sum_{N=0}^{\infty}\left\{\frac{(N+3)!}{2N!}
   {f^0}^N+\sum_{n=1}A_{7}(N,n){f^0}^{N-2n}(ff)^n\right\}\\
& &\times\xi^{mn}\left(\xi_{(ij)}f_{(mn}f_{kl)}+\xi_{(kl)}f_{(mn}f_{ij)}\right)\\
& &-\frac{8}{9\lambda^2}\sum_{N=0}^{\infty}\left\{(N+1)
   {f^0}^N+\sum_{n=1}A_{8}(N,n){f^0}^{N-2n}(ff)^n\right\}\\
& &\times\xi_{(mn)}\xi_{(kl)}\\
& &-\frac{4}{135\lambda^2}\sum_{N=0}^{\infty}\left\{\frac{(N+1)!}{12N!}
   {f^0}^N+\sum_{n=1}A_{9}(N,n){f^0}^{N-2n}(ff)^n\right\}\\
& &\times(\xi f)^{2}f_{(ij}f_{kl)}\\
& &-\frac{1}{945\lambda^2}\sum_{N=0}^{\infty}\left\{\frac{(N+7)!}{36N!}
   {f^0}^N+\sum_{n=1}A_{10}(N,n){f^0}^{N-2n}(ff)^n\right\}\\
& &\times\left((\xi^f)+\xi^{mn}f^{rs}f_{(mn}f_{rs}f_{ij}f_{kl)}\right)\\
& &+\frac{1}{693\lambda^2}\sum_{N=0}^{\infty}\left\{\frac{(N+7)!}{240N!}
   {f^0}^N+\sum_{n=1}A_{11}(N,n){f^0}^{N-2n}(ff)^n\right\}\\
& &\times\xi^{he}\xi^{mn}f^{rs}f_{(he}f_{mn}f_{rs}f_{ij}f_{kl)}\\
& &-\frac{1}{105\lambda^2}\sum_{N=0}^{\infty}\left\{\frac{(N+5)!}{12N!}
   {f^0}^N+\sum_{n=1}A_{12}(N,n){f^0}^{N-2n}(ff)^n\right\}\\
& &\times\left\{\xi^{mn}f^{rs}\left(f_{(mn}f_{rs}f_{kl)}\xi_{(ij)}
   +f_{(mn}f_{rs}f_{ij)}\xi_{(kl)}\right)+
   2(\xi^ f)\xi^{mn}f_{(mn}f_{ij}f_{kl)}\right\}\\
& &+\frac{2}{27\lambda^2}\sum_{N=0}^{\infty}\left\{\frac{(N+3)!}{N!}
   {f^0}^N+\sum_{n=1}A_{13}(N,n){f^0}^{N-2n}(ff)^n\right\}\\
& &\times (\xi^ f) \left(\xi_{(ij)}f_{(kl)}+\xi_{(kl)}f_{(ij)}\right)\\
\end{eqnarray*}

\begin{eqnarray*}
B_{ij}
&=&-\frac{4}{3\lambda^2}\sum_{N=0}^{\infty}\left\{\frac{(N+2)!}{N!}
  {f^0}^N+\sum_{n=1}B_{1}(N,n){f^0}^{N-2n}(ff)^n\right\}f_{(ij)}\\
& &+\frac{8}{\lambda^2}\sum_{N=0}^{\infty}\left\{{f^0}^N
   +\sum_{n=1}B_{2}(N,n){f^0}^{N-2n}(ff)^n\right\}\frac{1}{f^{(ij)}}\\
& &+\frac{1}{15\lambda^2}\sum_{N=0}^{\infty}\left\{\frac{(N+3)!}{N!}
  {f^0}^N+\sum_{n=1}B_{3}(N,n){f^0}^{N-2n}(ff)^n\right\}\xi^{mn}f_{(mn}f_{ij)}\\
& &+\frac{8}{3\lambda^2}\sum_{N=0}^{\infty}\left\{{(N+1)}
  {f^0}^N+\sum_{n=1}B_{4}(N,n){f^0}^{N-2n}(ff)^n\right\}\xi_{(ij)}\\
& &+\frac{1}{315\lambda^2}\sum_{N=0}^{\infty}\left\{\frac{(N+8)!}{6!N!}
  {f^0}^N+\sum_{n=1}B_{5}(N,n){f^0}^{N-2n}(ff)^n\right\}\\
& &\times\xi^{eh}\xi^{mn}f^{rs}f^{tv}f_{(eh}f_{mn}f_{rs}f_{tv}f_{ij)}\\
& &+\frac{2}{27\lambda^2}\sum_{N=0}^{\infty}\left\{\frac{(N+4)!}{3N!}
  {f^0}^N+\sum_{n=1}B_{6}(N,n){f^0}^{N-2n}(ff)^n\right\}(\xi f)(\xi f)f_{ij)}\\
& &+\frac{1}{105\lambda^2}\sum_{N=0}^{\infty}\left\{\frac{(N+6)!}{18N!}
  {f^0}^N+\sum_{n=1}B_{7}(N,n){f^0}^{N-2n}(ff)^n\right\}(\xi f)\xi^{mn}f^{rs}
  f_{(mn}f_{rs}f_{ij)}\\
& &-\frac{1}{315\lambda^2}\sum_{N=0}^{\infty}\left\{\frac{(N+6)!}{48N!}
  {f^0}^N+\sum_{n=1}B_{8}(N,n){f^0}^{N-2n}(ff)^n\right\}\xi^{eh}\xi^{mn}f^{rs}
  f_{(eh}f_{mn}f_{rs}f_{ij)}\\
& &-\frac{1}{45\lambda^2}\sum_{N=0}^{\infty}\left\{\frac{(N+4)!}{2N!}
  {f^0}^N+\sum_{n=1}B_{9}(N,n){f^0}^{N-2n}(ff)^n\right\}\\
& &\times(\xi_{ij}\xi^{(mn}f^{rs)}f_{(mn}f_{rs)}+(\xi f)\xi^{mn}f_{(ij}f_{mn)})\\
& &-\frac{4}{9\lambda^2}\sum_{N=0}^{\infty}\left\{\frac{(N+2)!}{N!}
  {f^0}^N+\sum_{n=1}B_{10}(N,n){f^0}^{N-2n}(ff)^n\right\}(\xi f)\xi_{(ij)}\\
\end{eqnarray*}
\begin{eqnarray*}
C_{ij}
&=&-\frac{1}{30\lambda^2}\sum_{N=0}^{\infty}\left\{\frac{(N+2)!}{2N!}
  {f^0}^N+\sum_{n=1}C_{1}(N,n){f^0}^{N-2n}(ff)^n\right\}\xi^{mn}f_{(mn}f_{ij)}\\
& &-\frac{8}{3\lambda^2}\sum_{N=0}^{\infty}\left\{{f^0}^N+\sum_{n=1}C_{2}(N,n){f^0}^
  {N-2n}(ff)^n\right\}\xi_{(ij)}\\
& &+\frac{1}{35\lambda^2}\sum_{N=0}^{\infty}\left\{{f^0}^N+\sum_{n=1}C_{3}
   (N,n){f^0}^{N-2n}(ff)^n\right\}\xi^{mn}f^{rs}f_{(mn}f_{rs}f_{ij)}\\
& &+\frac{1}{9\lambda^2}\sum_{N=0}^{\infty}\left\{{f^0}^N+\sum_{n=1}C_{4}
   (N,n){f^0}^{N-2n}(ff)^n\right\}(\xi f)f_{(ij)}\\
\end{eqnarray*}
\begin{eqnarray*}
D
&=&-\frac{1}{15\lambda^2}\sum_{N=0}^{\infty}\left\{\frac{(N+3)!}{N!}
 {f^0}^N+\sum_{n=1}D_{1}(N,n){f^0}^{N-2n}(ff)^n\right\}\xi^{(ij}f^{kl)}f_{(ij}f_{kl)}\\
& &-\frac{8}{3\lambda^2}\sum_{N=0}^{\infty}\left\{(N+1){f^0}^N+\sum_{n=1}D_{2}(N,n){f^0}^{N-2n}(ff)^n\right\}(\xi f)\\
\end{eqnarray*}
\begin{eqnarray*}
E
&=&-\frac{8}{3\lambda^2}\sum_{N=0}^{\infty}\left\{(N+3){f^0}^N+\sum_{n=1}E_{1}(N,n){f^0}^{N-2n}(ff)^n\right\}\\
& &+\frac{8}{9\lambda^2}\sum_{N=0}^{\infty}\left\{\frac{(N+3)!}{6N!}{f^0}^N+\sum_{n=1}E_{2}(N,n){f^0}^{N-2n}(ff)^n\right\}(\xi f)^2\\
& &+\frac{4}{45\lambda^2}\sum_{N=0}^{\infty}\left\{\frac{(N+5)!}
 {12N!}{f^0}^N+\sum_{n=1}E_{3}(N,n){f^0}^{N-2n}(ff)^n\right\}(\xi f)\xi^{(ij}f^{kl)}f_{(ij}f_{kl)}\\
\end{eqnarray*}
\begin{eqnarray*}
F
&=&+\frac{8}{\lambda^2}\sum_{N=0}^{\infty}\left\{(N+1){f^0}^N+\sum_{n=1}
   F_{1}(N,n){f^0}^{N-2n}(ff)^n\right\}\\
& &-\frac{1}{63\lambda^2}\sum_{N=0}^{\infty}\left\{\frac{(N+7)!}{6!N!}{f^0}^N
   +\sum_{n=1}F_{2}(N,n){f^0}^{N-2n}(ff)^n\right\}\xi^{(ij}
   \xi^{kl}f^{mn}f^{rs)}f_{(ij}f_{kl}f_{mn}f_{rs)}\\
& &-\frac{1}{45\lambda^2}\sum_{N=0}^{\infty}\left\{\frac{(N+5)!}{3N!}{f^0}^N+
   \sum_{n=1}F_{3}(N,n){f^0}^{N-2n}(ff)^n\right\}(\xi f)\xi^{(ij}f^{kl)}
   f_{(ij}f_{kl)}\\
& &-\frac{4}{5\lambda^2}\sum_{N=0}^{\infty}\left\{\frac{(N+3)!}{3N!}{f^0}^N
   +\sum_{n=1}F_{4}(N,n){f^0}^{N-2n}(ff)^n\right\}(\xi f)^2\\
\end{eqnarray*}

We precise here below the convention notations used in writing our derived metric.

a) In writing the formulas, we have used the following notation
\begin{eqnarray*}
(f.f)= f^{(nm)}f_{(nm)}\\
\end{eqnarray*}

b) The coefficients $ A_{i}(N,n),B_{i}(N,n),...F_{i}(N,n)$ are finite
numerical values defined for $n\geq1$  and $N\geq 2n$. Simple examples are given
by
\begin{eqnarray*}
A_{1}(2,1)&=&-\frac{5}{6}\\
B_{1}(2,1)&=&-\frac{4}{6}\\
C_{1}(2,1)&=&-\frac{1}{6}\\
D_{1}(2,1)&=&-\frac{5}{6}\\
E_{1}(2,1)&=&-\frac{3}{6}\\
F_{1}(2,1)&=&-\frac{3}{6}\\
&\vdots&
\end{eqnarray*}

c) One can easily read this metric by remarking the fact that in each
term of the components $A_{ijkl}, B_{ij}, C_{ij}, D, E$ and $F$, we
have a general coefficient term given by
\begin{eqnarray*}
K(\star, N,n,f) =(\alpha(\star)){f^0}^{N}+\sum_{n=1}(\beta(\star)){f^0}^{N-2n}(ff)^n
\end{eqnarray*}
where $(\star)$ and $\alpha(\star)$, $\beta(\star)$ denote
respectively, the components $A_{ijkl}, B_{ij}, C_{ij}, D, E, F$, and their associate coefficients in the direction of ${f^0}^N$ and
${f^0}^{N-2n}(ff)^{n}$. As an example we consider
\begin{eqnarray*}
K(A_{ijkl}, N,n,f) =(\frac{N+3}{2N!}){f^0}^{N}+\sum_{n=1} 
A_{1}(N,n){f^0}^{N-2n}(ff)^n
\end{eqnarray*}
We guess that the terms $K(\star, N,n,f)$ described previously, should
describe some general term of mathematical series which can be very
useful to simplify more the obtained metric.

d) We point out that a simple choice can be done for $\lambda=1$ at the
level of Eq.(3.20) which reduce to
\begin{eqnarray*}
S=&\int &d^{2}zdu \partial_{++}\partial_{--}\omega
\{8-2\lambda{\xi^{--}}^{2}{\partial^{++}}^{2}\omega \}
\end{eqnarray*}
This choice correspond simply to cancel in Eq.(4.2) the terms
proportional to $\xi^{--}$.

\newpage


\begin{references}
\bibitem{1} A. Galperin, E. Ivanov, V. Ogievetsky and E. Sokatchev,
        JETP Lett. 40 (1984) 912

    A. Galperin, E. Ivanov, S. Lalitzin, V. Ogievetsky and E. Sokatchev,
        Class. Quant. Grav. 1 (1984) 469

\bibitem{2} A. Galperin, E. Ivanov, V. Ogievetsky and P.K. Townsend
        Class. Quant. Grav. 3 (1984) 625

    A. Galperin, E. Ivanov, V. Ogievetsky and E. Sokatchev,
       Comm. Math. Phys. 103 (1986) 515-526

\bibitem{3} L. Alvarez Gaum\'e and D.Z. Freedman,
       Comm. Math. Phys. 80 (1981) 443

\bibitem{4} N.J. Hitchin, A. Karlhede, U. Lindstrom and M. Rocek,
       Comm. Math. Phys. 108 (1987) 535

\bibitem{5} G.W. Gibbons and N.S. Manton, Phys. Lett. B 356 (1995) 32

    G.W. Gibbons and P. Rychenkova, hep-th/9608085

\bibitem{6} M. Atiyah and N.Hitchin, Phys. Lett. A107(1985)21
		
	G. Gibbons and N. Manton, Nucl. Phys. B274(1986)183
\bibitem{7} P. How, K. Stelle and P.K. Townsend
        Nucl. Phys. B214 (1983) 319

\bibitem{8} A. Galperin, E. Ivanov, V. Ogievetsky
         Nucl. Phys. B214 (1987) 74

\bibitem{9} U U. Lindstrom, M. Rocek
         Nucl. Phys. B222 (1983) 285

\bibitem{10} E.H. Saidi and M.B. Sedra
         Mod. Phys. Lett. AV9N34 (1994) 3163

\bibitem{11} A. Leznov and M. Saviliev, Lett. Math. Phys. 3(1979) 489,
	Comm. Math. Phys. 74(1980)111
	P. Mansfield, Nucl.Phys. B208 (1982) 277.

\bibitem{12} E. Witten, hep-th/9703166
\end{references}
\end{document}